\documentclass[preprint,12pt]{elsarticle}

\let\today\relax
\makeatletter
\def\ps@pprintTitle{%
    \let\@oddhead\@empty
    \let\@evenhead\@empty
    \def\@oddfoot{\footnotesize\itshape
         { } \hfill\today}%
    \let\@evenfoot\@oddfoot
    }
\makeatother

\usepackage{amssymb}
\usepackage{amsmath}

\usepackage{siunitx}
\usepackage{xcolor}

 \usepackage{lineno}

\begin{document}
\date{}

\begin{frontmatter}



\title{Molecular Weight-Dependent Evaporation Dynamics and Morphology of PEG Sessile Drops on Hydrophobic Substrates} 


\author[label1]{Feiyu An} 
\author[label1]{Junyi Ye} 
\author[label1]{Huanshu Tan\corref{cor1}} 

\cortext[cor1]{tanhs@sustech.edu.cn}
\affiliation[label1]{organization={Multicomponent Fluids group, Center for Complex Flows and Soft Matter Research \& Department of Mechanics and Aerospace Engineering, Southern University of Science and Technology},
            city={Shenzhen},
            postcode={518055}, 
            country={P.R. China}}

\begin{abstract}
The evaporation dynamics of sessile drops are crucial for material deposition in applications like inkjet printing and pharmaceutical development. 
However, the evaporation behavior of high molecular weight polymer solutions and their impact on deposit morphology and flow fields are not well understood. 
This study investigates the evaporation dynamics and deposit morphology of polyethylene glycol (PEG) solution drops on hydrophobic substrates, with molecular weights ranging from \num{200} to \num{1000}k \unit{\gram\per\mole}, covering five orders of magnitude. 
The results show that vapor diffusion dominates the evaporation process across all PEG molecular weights. 
Using image analysis and micro-particle image velocimetry ($\mu$-PIV), we reveal that molecular weight affects contact line dynamics and internal flow, leading to diverse deposit morphologies, including spherical caps, pillars, pool-shaped disks, and flat disks. 
Transient divergence and Péclet number calculations further confirm the role of hydrodynamics in deposit formation. 
These findings provide insights into the hydrodynamic and thermodynamic factors governing evaporation in polymeric sessile drops, with implications for material fabrication and the development of inkjet printing and coating techniques.
\end{abstract}




\end{frontmatter}



\section{Introduction}
\label{Introduction}

A sessile drop, derived from the Latin \textit{sessilis} meaning ``sitting'', refers to a solution drop resting on a surface.
In 1997, Deegan and colleagues made a groundbreaking discovery by demonstrating how interior capillary flows in a drying coffee drop shape its residual stain pattern, now famously known as the ``coffee-ring-stain'' effect~(\citet{Deegan1997,larson2017twenty}). 
This work highlighted the potential to control residual patterns~(\citet{deegan2000contact,hu2002evaporation,Popov2005,Cazabat2010,sefiane2014patterns,zang2019evaporation}) formed after the evaporation of microliter-scale solutions on surfaces, with applications in healthcare diagnostics~(\citet{sobac2011structural}), surface patterning~(\citet{feng2018droplet}), inkjet printing~(\citet{lohse2022fundamental}), and semiconductor fabrication~(\citet{maenosono1999growth,diao2014morphology}).
Since then, research on evaporating sessile drops has flourished.
Studies have explored phenomena such as moving contact line dynamics~(\citet{murisic2011evaporation,Stick-Slide,Stick-Jump}),  internal flow structures~(\citet{barmi2014convective}), the impact of surface hydrophobicity~(\citet{Gelderblom2011}) and micro/nanostructures~(\citet{marin2012building}), particle-laden drop deposit patterns~(\citet{yunker2011suppression}), and models that predict evaporation rates~(\citet{Popov2005,brutin2022drying,Stick-Slide,Stauber2014onthelifetimes,stauber2015a}).

Studies on everyday drying sessile drops, such as Whiskey~(\citet{kim2016controlled}) and Ouzo drops~(\citet{Tan2016,Tan2017}), have revealed dynamic surfactant and solutal Marangoni flows that develop within complex liquid drops during evaporation.
Along with dynamic interfacial flows, the preferential evaporation of volatile components triggers composition-driven dynamics, including liquid-liquid phase separation~(\citet{Tan2016,diddens2017evaporating}), segregation~(\citet{li2018evaporation}), and particle migration~(\citet{may2022phase}).
These multicomponent sessile drops are observed in prebiotic compartmentalization within coacervate systems as well~(\citet{guo2021non,qi2024multicompartmental}) and hold potential for advanced material fabrication~(\citet{moerman2018emulsion,Tan2019}). 
This growing body of research~(\citet{lohse2020physicochemical,wang2022wetting,gelderblom2022evaporation,tan2023self}) highlights the importance of understanding how individual components influence the evaporation dynamics of sessile drops.

Among the diverse range of multicomponent systems, polymer solutions provide a unique platform to investigate evaporation-driven transport and deposition phenomena due to their tunable physicochemical properties.
Polyethylene glycol (PEG), in particular, stands out for its exceptional biocompatibility, stability, and water solubility, making it a versatile material available in various chain lengths and molecular weights.
In dilute PEG-water solutions, surface tension decreases as concentration  increases~(\citet{ST2}), while viscosity rises exponentially, with higher molecular weights leading to greater viscosity~(\citet{ST5}).
Notably, PEG remains in a liquid phase at low molecular weights but transitions to a solid phase at higher molecular weights in ambient conditions~(\citet{PEGVaporPressure}). 
These distinct properties make the PEG-water sessile drops an ideal system for systematically investigating the molecular-level effects on the evaporation dynamics of sessile drops.

Given the widespread applications of macromolecule sessile drops in areas such as inkjet printing~(\citet{de2004inkjet,lohse2022fundamental}) and pharmaceutical development~(\citet{kim2012designing,fang2006drying}), significant interests has been directed toward understanding the evaporation behavior of macromolecule solutions and the deposits patterns they form on surfaces after evaporation~(\citet{zang2019evaporation}).
Research has shown that polymer sessile drops exhibit complex drying dynamics influenced by factors such as contact line pinning, drop size, radial and Marangoni flows, viscosity changes, and interactions with added salts~(\citet{Pauchard_2003,pauchard2003stable,de2004inkjet,fukai2006effects}. 
These processes result in diverse deposition patterns, including dot-like features, concentric rings, central buckling structures like the ``Mexican hat,'' and fractal or dendritic patterns when ions are present~(\citet{pauchard2003stable,willmer2010growth,giri2013multifractal,dutta2013experiment}).
While studies have demonstrated that molecular weight affects evaporation dynamics and deposit morphology -- such as increased pinning effects~(\citet{Pauchard_2003,baldwin2015classifying,PolymerMode2017,SongPolymer2024}) and Péclet number related changes in final deposit structures~(\citet{baldwin2012monolith,BALDWIN2014}) -- the mechanisms underlying these phenomena remain poorly understood.
In particular, the evolution of internal flows as a function of molecular weight has yet to be comprehensively examined.

To fill these gaps, this study explores the impact of polymer molecular weight on the evaporation dynamics, internal flow patterns, and final deposit morphology of PEG solution droplets on hydrophobic substrates. 
By systematically varying molecular weights and initial concentrations, we measure droplet profiles and characteristic parameters to reveal how molecular weight influences evaporation behavior. 
Micro-Particle Image Velocimetry ($\mu$-PIV) is employed to track the evolution of internal flow fields, providing detailed insights into the hydrodynamic processes involved. 
This study aims to improve our understanding of polymer droplet evaporation and its applications in material fabrication.

\section{Materials and methods}
\label{Materials and methods}
\subsection{Sessile Drop Solutions}
\label{PEG drop solution}
Evaporating drop of polyethylene glycol (PEG) solutions was used as model systems to investigate the effects of soluble polymer additives on the evaporation behavior of sessile drops.
PEGs with average molecular weights of \SI{200}{\gram\per\mole}, \SI{8000}{\gram\per\mole}, and \SI{20000}{\gram\per\mole} are interchangeably listed as PEG 200, PEG 8k, and, PEG 20k, respectively.
PEG solutions were prepared by mixing ultra-pure water with PEG polymers of various molecular weights, including 200, 1k, 6k, 8k, 10k, 20k, 70k, 100k, 300k, and 1000k g/mol (purchased from Macklin) , at mass fractions ranging from 0.5\% (w/w) to 16\% (w/w).
All chemicals were used without further purification.
To prevent damage to the polymer molecular chains, all solutions were gently mixed using a roller mixer (JOANLAB, RM-6pro) with a speed of \SI{30}{rpm}.

\subsection{Preparation of Hydrophobic Substrates}
\label{substrate preparation}
Hydrophobic substrates were fabricated by coating microscope glass slides (Matsunami, S7224) and cover glass (Matsunami, NEO) with a layer of 1H,1H,2H,2H-perfluorodecyltriethoxysilane (PFDS).
The preparation process is detailed as follows.
First, the glass slides were carefully wiped with ethanol-wetted tissue to mechanically remove surface contaminants.
The slides were then sonicated sequentially in fresh acetone (YONGHUA, $\ge$\SI{99.5}{\percent}, AR), ethanol (General Reagent, $\ge$\SI{99.7}{\percent}, AR), and Milli-Q water, each for 15 minutes, to eliminate organic residues and any remaining detergent.
After sonication, the slides were dried with compressed nitrogen and further cleaned in a plasma cleaner (HARRICK PLASMA, PDC-32G-2) for 10 minutes~(\citet{Tan2019}).
Following cleaning, the glass slides were immersed in a coating solution containing \SI{1}{vol\percent} 1H,1H,2H,2H-perfluorodecyltriethoxysilane (PFDS, Macklin, \SI{96}{\percent}) and \SI{99}{vol\percent} ethanol at \SI{60}{\celsius} for 4 hours.
The coated slides were then dried with compressed nitrogen and placed in a clean Petri dish at \SI{60}{\celsius} for an additional 4 hours to remove any unbound PFDS.
Finally, the slides were carefully stored in the Petri dish to avoid any contaminants in the air.
The resulting hydrophobic glass slides exhibited advancing and receding contact angles of \SI{107}{\degree}  and \SI{95}{\degree}, respectively.

\subsection{Experimental Setup}
\label{SetUp}
\begin{figure}[t]
  \centering
    \includegraphics[width=0.9\columnwidth]{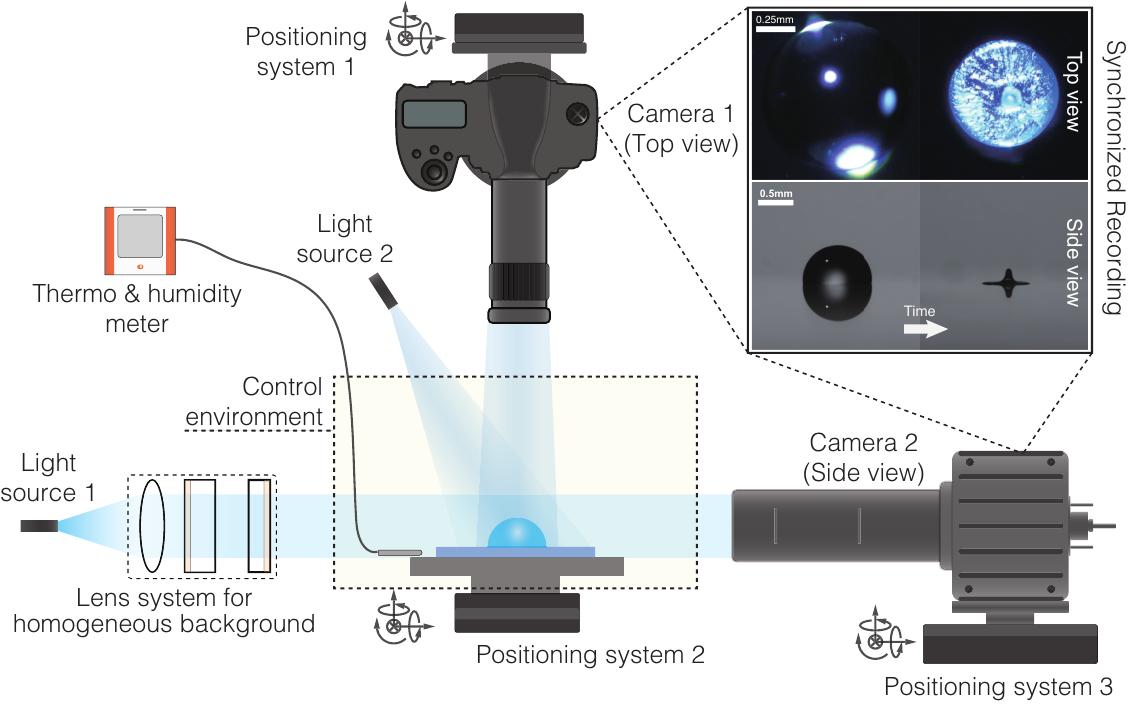}
    \caption{Experimental set-up showing the evaporation of an PEG solution drop being recorded by two synchronized cameras. 
    }
    \label{setup}
\end{figure}
Drop evaporation experiments were conducted in a custom-built chamber where environmental conditions, including temperature and humidity, were controlled.
As shown in Figure~\ref{setup}, a \SI{0.5}{\mu \liter} PEG drop was deposited onto a hydrophobic glass substrate using a syringe pump (LEADFLUID, TYD01). 
The entire evaporation process was recorded using two cameras.
One camera [Ximea; MD091MU-SY] captured side-view images at 1 frame per second (fps) with a resolution of 1688 $\times$ 1352 pixels, paired with a telecentric lens [Edmund; 67306].
The other camera [Nikon; Z8] recorded top-view images at 24 frames per second (fps) with a resolution of 1920 $\times$ 1080 pixels, fitted with a high-magnification zoom lens system [Thorlabs; MVL12X12Z]. 
A custom-built collimated light source provided a uniform background for side-view imaging, and a separate LED light source was used to illuminate the top view, highlighting the dehydration process of the drying PEG drops. 
Temperature $T$ and relative humidity $RH$ near the evaporating drop were monitored every ten seconds using a temperature and humidity meter [Elitech; GSP-8A], with an accuracy of $\pm \SI{3}{\percent} RH$ within the range of \SI{20}{\percent} to \SI{80}{\percent} at \SI{25}{\celsius} and $\pm \SI{0.3}{\celsius}$ for temperature between $-20$ and $\SI{40}{\celsius}$. 
The probe was positioned around 5 cm from the drying drop. 

\subsection{$\mu$-PIV Measurements}
\label{met:muPIV}
Micro-particle imaging velocimetry ($\mu$-PIV) were used to capture the flow field during the evaporation of the PEG sessile drop.
To track the flow, a suspension of \SI{4}{\micro\liter} polystyrene (PS) microspheres (YuanBiotech, \SI{2}{\micro\meter} diameter, with \SI{10}{\milli\gram\per\milli\liter} solid content) was added to \SI{0.5}{\milli\liter} of the PEG solution. All particles were cleaned by diluting with ultrapure water, \SI{5}{\minute} sonication, then \SI{5}{\minute} centrifugation and followed by replacing the water (dispersing agent) with fresh water. This process was repeated for 5 times to eliminate the surfactants in the suspension of PS particles.
To verify the accuracy of the tracer particles, the Stokes number was calculated as $St=t_0 u_{max}/R_0$, where $t_0=(\rho_p d_p^2/18\mu)(1+\rho/(2\rho_p))$ represents the characteristic response time.
In this equation, $u_{max}$ is the maximum fluid velocity ($\sim$\SI{5}{\micro\meter/\second}), $R_0$ is the initial radius of contact area ($\sim$\SI{1}{\milli\meter}), $\rho_p$ is the density of tracer particles (\SI{1.05e3}{\kilogram\per\cubic\meter}), $\rho$ is the density of the aqueous PEG solution (about \num{1.0}$\sim$\SI{1.12e3}{\kilogram\per\cubic\meter})\cite{zivkovic2013volumetric}, $d_p$ is the diameter of particles (\SI{2}{\micro\meter}), and $\mu$ is the dynamic viscosity of the solution (\num{e-3}$\sim$\SI{e3}{\pascal\second})\cite{ST3,ST5}.
The resulting Stokes number $St\approx$\num{e-9}$\sim$\num{e-15}, confirming that the particles effectively traced the liquid flow inside the evaporating drops.
The $\mu$-PIV experiments were conducted on a custom-built setup, with bottom-view and side-view images captured by two high-speed cameras [Photron; Mini-Ax200, 1024 pixel $\times$ 1024 pixel resolution] simultaneously. For illumination, we took advantage of two laser sheets with the same wavelength of 532 nm. One was placed horizontally to illuminate the bottom layer near the glass substrate for bottom-view imaging. The other was in the vertical direction for side-view imaging.
The acquired image sequences were analyzed using PIVLab software~(\citet{PIVlabThielicke_2021}), where each pair of consecutive images was processed to calculate flow fields. 
In the analysis, the interrogation area and step size were set to 64 pixels and 32 pixels, respectively, for the first pass, and 32 pixels and 16 pixels for the second pass.
Each pixel corresponds to a physical length of \SI{1.45}{\micro\meter} for bottom view and \SI{1.67}{\micro\meter} for side view.

\subsection{Image Analysis and Data Calculation}
\label{Image analysis}
Evaporation images data were analyzed using a custom MATLAB program. 
The initial contact line position of each drop was manually set for each data set.
In each side-view frame, the drop contour was approximated by fitting it to a circular shape.
To validate this approximation, we calculated the capillary length of the PEG solution, $\ell = \sqrt{\sigma/(\rho g)}$, where $\sigma$ is the surface tension of the aqueous PEG solution (about \SIrange{40}{70}{\milli\newton/\meter}) (\citet{ST1,ST2,ST3,ST4,ST5}).
The resulting capillary length, $\ell\approx$\SIrange{1.91}{2.67}{\milli\meter}, was larger than the size of the sessile drop, confirming that surface tension dominates over gravity, thereby justifying the circular fitting approach.
The contact angle $\theta$ and contact radius $R$ were calculated based on the intersection of the base line and the fitted circle in each frame. 
The drop volume $V$ was determined by integrating the areas of the profile, assuming rotational symmetry of the drop, a condition confirmed through top-view recordings.

\begin{figure}[ht!]
  \centering
  \includegraphics[width = \linewidth]{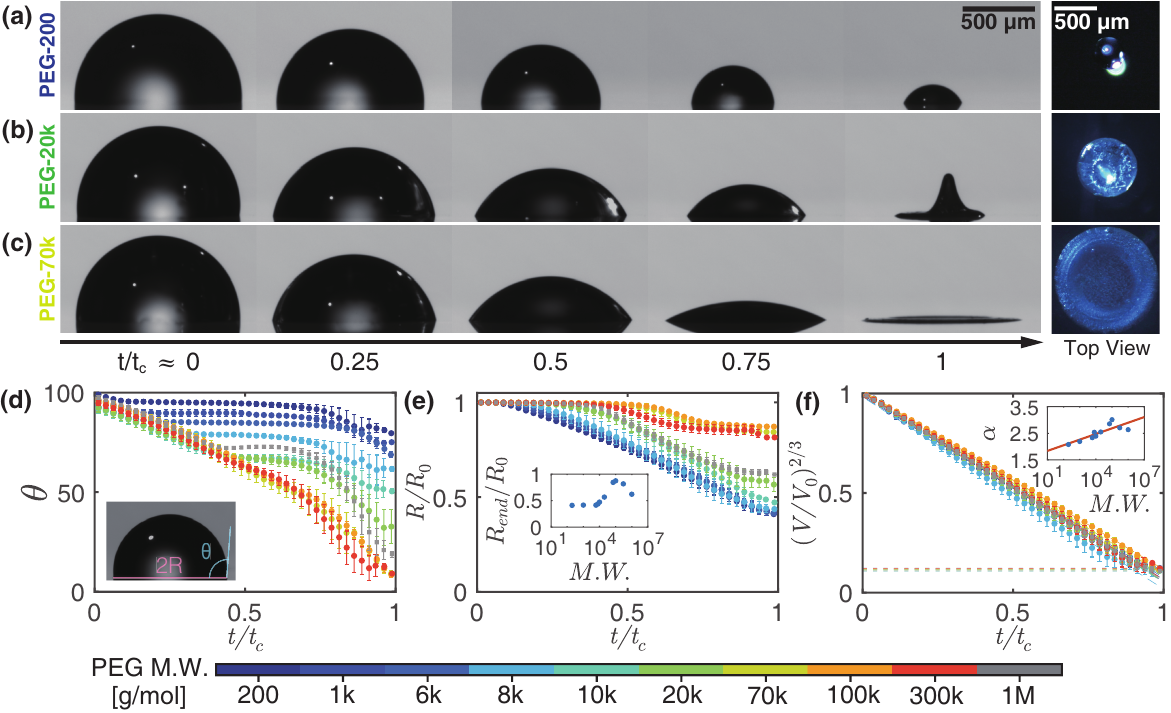}
  \caption{(a-c) Side view snapshots of evaporating PEG sessile drops (2\% w/w) on a hydrophobic substrate at five characteristic times for molecular weights of \SI{200}{\gram\per\mole}, \SI{20}{k\gram\per\mole}, and \SI{70}{k\gram\per\mole}. 
  The final column displays top views of the residual deposits after complete evaporation at $t/t_c =1$.
  Accompanying videos are available in Movies S1, S2, and S3.
  (d) Temporal evolution of the contact angle ($\theta$) for PEG additives with molecular weights ranging from 200 g/mol to 1000k g/mol. 
  The inset defines $R$ and $\theta$.
  The initial PEG concentration of the drops is 2\% (w/w). 
  (e) The corresponding temporal evolution of the normalized lateral size ($R/R_0$), where $R_0$ indicates its initial value, across varying molecular weights of PEG. 
  The inset shows the dependence of the final contact radii of deposits on the PEG molecular weight.
(f) Shows the linear relationship between $(V/V_0)^{2/3}$ and time $t$, illustrating how the normalized drop volume decreases over time. 
The inset illustrates how the evaporation rate slope $\alpha$ varies with PEG molecular weight.
  The color of each line indicates the corresponding molecular weight of PEG additives. 
  Each dataset in panels (d-f) contains five data points, with error bars representing the standard deviation. The initial PEG concentration in the drops is consistently 2\% (w/w).
  The characteristic time $t_c$ is defined as $t_c = \frac{\rho R_0^2}{\alpha Dn_s(1-RH)}$.
  }
  \label{PEGdata}
 \end{figure}

\section{Results and Discussions}
\subsection{Polymer Molecular Weights Affect Drop Evaporation}
\label{Experimental results}

To examine effects of molecular weight on the evaporation of sessile PEG drops, we placed \SI{0.5}{\micro \liter} PEG solution drops of different molecular weights $MW$ on a flat hydrophobic substrate and recorded their evaporating process from synchronized side and top views.
The closed environment was controlled at $\num{20.3}\pm\SI{0.3}{\celsius}$ and $\SI{51}{}\pm\SI{5}{\percent}$ relative humidity.
The initial PEG concentrations of the drop solution are controlled to be \SI{2}{\percent} weight fractions in this section.

The side-view snapshots in Figures~\ref{PEGdata}(a-c) show three representative profile evolutions during the evaporation of PEG drops, demonstrating the influence of molecular weight on contact line dynamics.
The time $t/t_c$ is normalized by a diffusion-governed time scale $t_c=\frac{\rho R_0^2}{\alpha Dn_s(1-RH)}$, where $R_0$ is the initial contact radius, $\rho$ is the density of PEG solution (assuming $\rho = \rho_{water}$), $D$ is the vapor diffusivity, $n_s$ is the saturated vapor density, $RH$ is the relative humidity, and $\alpha$ is fitting parameter to ensure that the data curves asymptotically approach $t/t_c=1$.

For PEG with a molecular weight of $200$~\SI{}{\gram \per \mole} (Fig.~\ref{PEGdata}a), the drop predominantly evaporates during $t/t_c<0.75$ in a constant contact angle (CCA) mode, resulting in the formation of a spherical-cap deposit composed of precipitated PEG. 
At a molecular weight of $20$k~\SI{}{\gram \per \mole} (Fig.~\ref{PEGdata}b), the CCA period is shortened, and both the contact angle and contact area decrease throughout most of the evaporation, leading to the formation of a PEG pillar.
As the molecular weight increased to $70$k~\SI{}{\gram \per \mole} (Fig.~\ref{PEGdata}c), the drop exhibits a constant contact area with a continuously decreasing contact angle (CCR mode) for $t/t_c<0.75$, ultimately forming a flat PEG film disk.

The transition from CCA to CCR mode with increasing molecular weight is confirmed by a systematic study in which the polymer molecular weight is varied from $200$ to $1$k $6$k, $8$k, $10$k, $20$k, $70$k, $100$k, $300$k, and up to $1000$k~\SI{}{\gram \per \mole}, spanning five orders of magnitude.
To quantify this transition, we measured the evolutions of the contact angle $\theta$ (Fig.~\ref{PEGdata}d) and contact area radius $R$ (Fig.~\ref{PEGdata}e).
Molecular weights $M.W.$ are color-coded, and error bars represent standard deviations from five repeated measurements.
The contact radius $R$ and time $t$ are normalized by initial values $R_0$ and characteristic time $t_c$, respectively.
As PEG molecular weight increases, the flat region of the contact angle curve shortens, and the reduction in contact area radius becomes less pronounced.
The final contact area radius as a function of molecular weight $M.W.$ is shown in the inset of Figure~\ref{PEGdata}e.
The results indicate that when the molecular weight is small (from $\sim10^2$ to $\sim10^4$~\SI{}{\gram \per \mole}), molecular weight variation has little effect on the final size.
However, as the molecular weight increases, the final radius of the residuals increases.
Beyond a certain molecular weight,  around $\sim10^6$~\SI{}{\gram \per \mole}, further increases cause the final radius of the residuals to decrease again.

Although the variation in molecular weight affects the dynamics of moving contact line, the drop volume change adheres to the vapor-diffusion-controlled ``$D^2$'' law~(\citet{Cazabat2010}). 
As shown in Figure~\ref{PEGdata}f, the normalized drop volume decreases as $(V(t)/V_0)^{2/3} \sim t/t_c$, where $(V(t))^{2/3}$ approximates the effective surface area of the drop.
Since $t_c$ is dependent on the fitting parameter $\alpha$ ($t_c\sim \alpha^{-1}$), the relationship between $\alpha$ and the logarithm of the molecular weight ($M.W.$) is presented in the inset of Figure~\ref{PEGdata}f.
The observed monotonic increase of $\alpha \sim \log{MW}$
suggests that $t_c$ is a molecular weight-dependent characteristic time scale for evaporation, with PEG drops evaporating faster as the molecular weight increases.
Given that the saturation vapor pressure for PEG 200 is below \SI{0.05}{\pascal} and decreases further with increasing molecular weight~(\citet{PEGVaporPressure}), PEG molecules act as non-volatile solutes for varying molecular weights.. 
Therefore, the observed acceleration in evaporation is likely due to the transition to CCR mode, where a decreasing contact angle amplifies the evaporation flux singularity at the contact line(\citet{Cazabat2010,yunker2011suppression,Deegan1997,deegan2000contact}), thereby accelerating the process.

\subsection{Vapor-Diffusion Dominates Evaporation}

Since the evaporation of a PEG sessile drop is predominantly governed by vapor diffusion, to predict the mass loss rate of water $dM/dt$, we apply Popov's model~(\citet{Popov2005}), expressed as,
\begin{equation}
  \begin{aligned}
  \frac{dM}{dt}=&-\pi R_0D(n_s-n_{\infty})f(\theta),
  \end{aligned}
  \label{eq:Popov}
\end{equation}
where $R_0$ is the initial contact radius of the drop, $D$ is the diffusion coefficient of water vapor in air, $n_s$ is the saturated vapor density, and $n_\infty=RHn_s$ represents the ambient vapor density far from the drop, with $RH$ as the relative humidity.
The geometrical factor $f(\theta)$ encapsulates the influence of the drop shape and is defined as,
\begin{equation}
  \begin{aligned}
  f(\theta)=&\frac{\sin\theta}{1+\cos\theta}+~4\int_{0}^{\infty} \frac{1+\cosh2\theta\tau}{\sinh2\pi\tau}\tanh[(\pi-\theta)\tau] \, d\tau,
  \end{aligned}
  \label{eq:ftheta}
\end{equation}
where $\theta$ is the drop's contact angle.
While Popov's model accurately describes evaporation in the constant contact radius (CCR) mode~(\citet{Popov2005, Gelderblom2011,Tan2017,gelderblom2022evaporation,Stauber2014onthelifetimes,Stick-Slide}), it is not directly applicable to PEG drops, where evaporation often deviates from the CCR mode.

To address this, we generalized the model by replacing $R_0$ and $\theta$ with their time-dependent counterparts $R(t)$ and $\theta(t)$.
This extension applies Popov's predication over small time intervals where the variation in $R(t)$ and $\theta(t)$ remain negligible.
The generalized equation becomes,
\begin{equation}
  \frac{dM}{dt}=-\pi R(t)D(n_s-n_{\infty})f(\theta(t)),
  \label{pmt}
\end{equation}
Assuming the dilute PEG solution maintains a constant density ($\rho = \rho_{water}$), the total mass and time are normalized as~(\citet{Gelderblom2011})
\begin{equation}
  \hat{M}=\frac{M}{\rho R_{0}^{3}}=\frac{V}{R_{0}^{3}}, \quad   \hat{t}=\frac{2D(n_s-n_{\infty})t}{\rho (3V_0/(2\pi))^{2/3}},
  \label{Mhat}
\end{equation}
Substituting these nondimensional parameters into the generalized equation~(\ref{pmt}) yields,
\begin{equation}
  \frac{d\hat{M}}{d\hat{t}}=-\pi \frac{(3V_0/(2\pi))^{2/3}}{2R_0^3}R(\hat{t})f(\theta(\hat{t})).
  \label{mpm}
\end{equation}
This formulation works for PEG drop evaporation in both constant contact radius (CCR) and constant contact angle (CCA) modes, revealing that the mass loss rate depends dynamically on the contact radius and angle.

\begin{figure}[ht!]
\centering
  \includegraphics[width=0.6\linewidth]{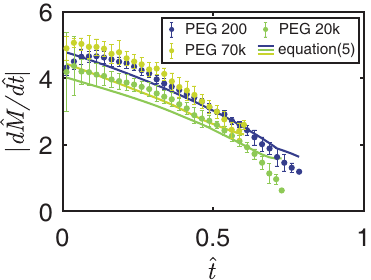}
  \caption{
   Dimensionless mass loss rates for evaporating PEG solution drops (2\% w/w) corresponding to Fig.2 (a-c). 
   The solid lines show predictions from the generalized Popov’s model (Eqn.~(5)), closely matching the experimental data scaled by Eqn.~(4). 
   Line colors correspond to different molecular weights of PEG, reflecting changes in contact motion modes from CCA to CCR.
  }
  \label{Popov model}
\end{figure}

Figure~\ref{Popov model} shows the evolution of mass loss rates for PEG drops with various molecular weights.
Experimental measurements are shown as data points, with error bars representing standard deviations from five repeated trials.
Solid lines correspond to theoretical predications based on the generalized model~(\ref{mpm}), where $R(t)$ and $\theta(t)$ are extracted from experimental data.
The results demonstrate good agreement between experiments and theoretical values for $t/t_c<0.7$.
Minor deviations occur in the final evaporation stage due to the breakdown of the dilute solution assumption. 
Overall, the generalized vapor-diffusion-based model captures the mass loss rate of sessile PEG drops across five orders of magnitude in molecular weight.

\begin{figure}[ht!]
\centering
  \includegraphics[width=\linewidth]{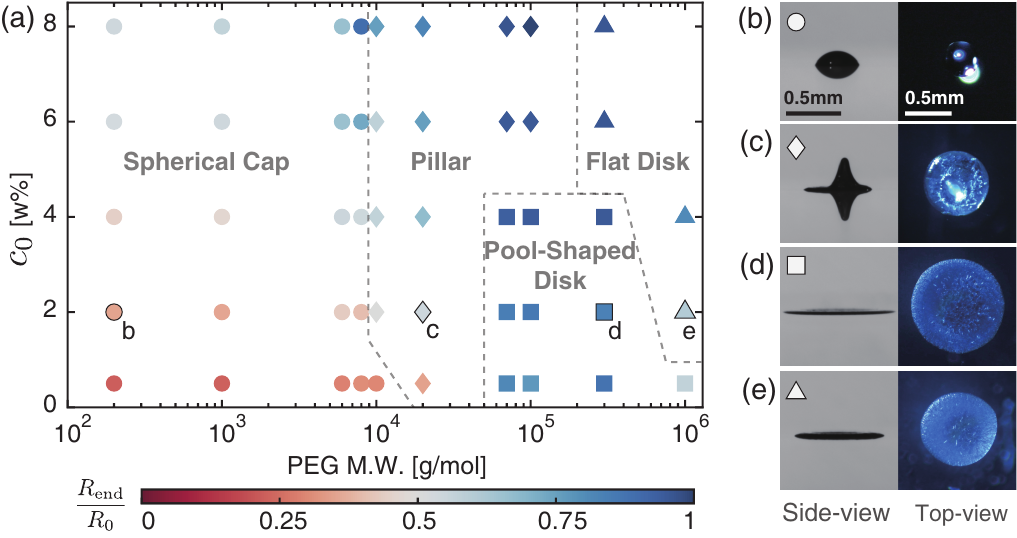}
  \caption{Deposit size and phase diagram of the final deposit morphology. (a) The deposit size versus molecular weight of PEG additives. The circle mark represents the deposit shrink to a spherical cap, while the diamond mark represents a deposit with pillar, the square mark represents a pool-shaped disk, and the triangle mark represents a flat disk. The colors of the data indicate the deposit size with respect to the initial drop size. (b-d) Experimental snapshots of the side-view perspective and the top-view perspective of the corresponding patterns of the final deposits (labels with margin in (a)). 
  }
  \label{deposit size}
\end{figure}

\subsection{Morphological Variations in Final Deposits}
\label{sec: deposit}
The molecular weight of PEG in a sessile drop also affects the morphology of the deposit residuals. 
Changes in molecular weight lead to distinct patterns, including spherical caps, pillar-like structures, pool-shaped disks, and flat disks.
Figure~\ref{deposit size} presents a deposit phase diagram developed by systematically varying the PEG molecular weight and initial PEG mass fraction.
Each symbol in the diagram represents a specific deposit pattern, with color coding indicating the final lateral size of the deposit $R_{end}$,  to the initial drop contact radius $R_0$.
These results were consistent across five independent experimental trials.

Lower molecular weight PEG solutions, such as \SI{200}{\gram \per \mole}, form spherical cap-like deposits upon evaporation (Fig.\ref{deposit size}b).
At the controlled temperature of $20.3\pm\SI{0.3}{\celsius}$, PEG-200 remains in the liquid phase~(\citet{PEGVaporPressure}) and behaves as a binary drop system with water~(\citet{li2018evaporation}).
Upon complete evaporation, the remaining PEG forms a spherical cap, with the contact angle corresponding to the equilibrium contact angle of a pure PEG-200 drop on the substrate ($\sim\SI{76}{\degree}$).
This explains the step change in contact angle from \SI{107}{\degree} to \SI{75}{\degree}, as shown in Figure~\ref{PEGdata}d.
Since no solid PEG-200 phase forms, the drop evaporates in CCA mode, shrinking smoothly with a relatively small final deposit size $R_{end}/R_0$.

With a slight increase in molecular weight, such as PEG-1000, PEG transitions to a solid phase at the same temperature~(\citet{PEGVaporPressure}).
Interestingly, after evaporation, the PEG-1000 residue remains in a metastable liquid state, preserving a spherical cap shape for days under controlled conditions.
A similar spherical cap structure forms in nanoparticle-landed Ouzo drops, which leads to the creation of supraparticles~(\citet{Tan2019,Vinayak2010supraparticle}).

As the molecular weight increases further, PEG-20k residuals evolve into pillar-like structures (Fig.\ref{deposit size}c), a phenomenon also observed in previous studies~(\citet{baldwin2012monolith,BALDWIN2014,Baldwin2011}).
These studies highlight the importance of a Péclet-type number~(\citet{BALDWIN2014}), which quantifies the balance between evaporation-flux-driven advection and the diffusive motion of the dissolved polymer.
However, this time-independent number overlooks the complex transition between contact line moving modes and internal bulk flows. 
The dynamics of these motions, along with the evolution of the corresponding classical Péclet number -- describing the ratio of advective and diffusive motion -- remain poorly understood.

As the molecular weight of PEG increases to approximately 100k, the residuals shift from pillar-like 3D structures to pool-shaped 2D films (Fig.\ref{deposit size}d).
The drops primarily evaporate in a CCR mode, as shown in Figures~\ref{PEGdata}d and \ref{PEGdata}e.
The reduction in contact angle increases the evaporation flux at the contact line~(\citet{Cazabat2010, Popov2005}), which promotes the preferential precipitation of solid PEG near the contact edge~(\citet{Tan2016}).
As PEG at higher molecular weights is in the solid phase~(\citet{PEGVaporPressure}),
the precipitated PEG molecules at the contact line enhances the pinning effect, further amplifying capillary flow~(\citet{Deegan1997}).
Thus, the lateral size of the deposit is large with $R_{end}/R_0 \approx 1$.
This results in more PEG accumulation at the contact line and the formation of a pool-shaped 2D film.

At sufficiently high molecular weight, such as $1000$k~\SI{}{\gram \per \mole}, PEG forms flat, disk-like residuals (Fig.\ref{deposit size}e).
The evaporation process follows a sequence from CCR to CCA mode and back to CCR, as shown in Figures~\ref{PEGdata}d and \ref{PEGdata}e.
The CCA mode reduces the final deposit's contact area, reflecting complex contact line dynamics, though the exact mechanism remains unclear.
After evaporation, the solidified PEG disks often deform and detach from the substrate, likely due to stronger intermolecular interactions at higher molecular weights.

The progression of residual morphologies -- from spherical caps to pillars, and finally disks -- mainly follows the increase in PEG molecular weight, although varying the initial solution concentration also influences the deposit pattern (Fig.~\ref{PEGdata}a).
Notably, a pool-shaped disk phase appears at lower initial concentrations.
These distinct morphologies are linked to different contact line dynamics.
However, the effects of hydrodynamics on these morphologies remain to be explored.

\begin{figure}[ht!]
 \centering
 \includegraphics[width = \linewidth]{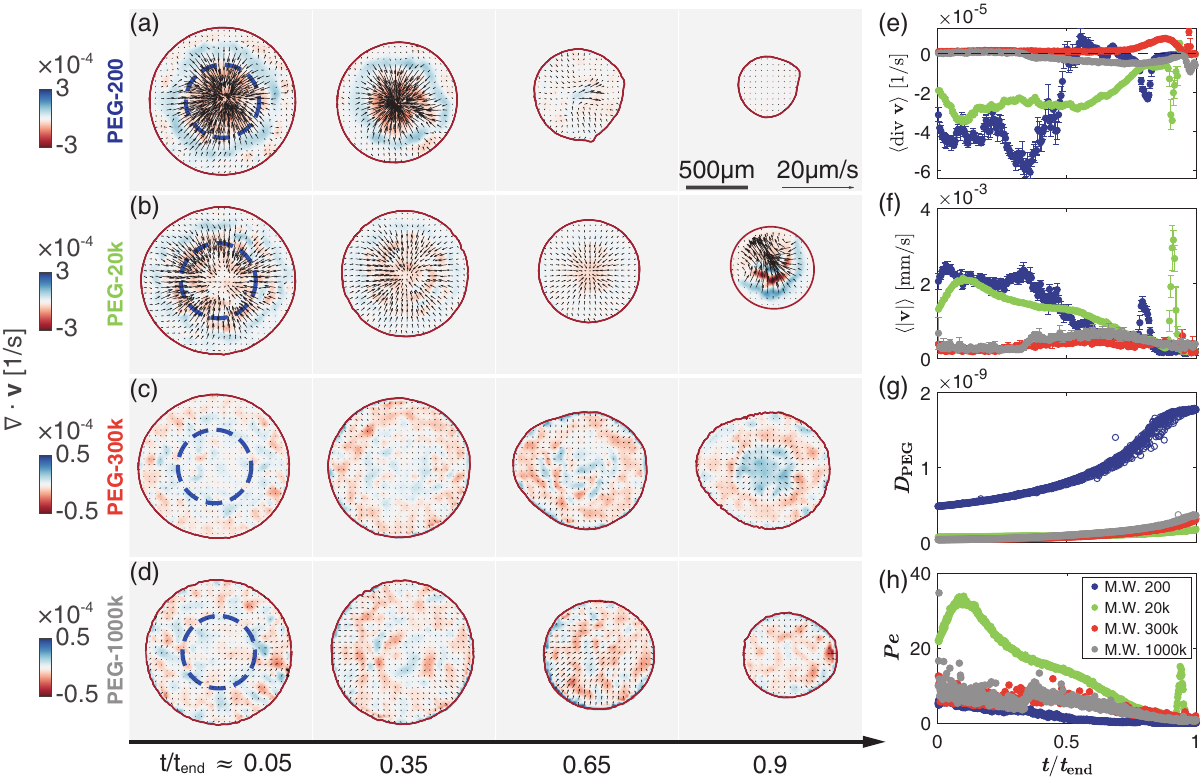}
 \caption{Visualization of the divergence and flow field variations in evaporating drops, forming structures of (a) spherical cap, (b) pillar, (c) pool-shaped disk, and (d) flat disk morphologies. 
The transient drop contact line are marked by red lines, while black arrows depict the local velocity fields at corresponding times. 
The divergence within the drops is color-coded, with intensities specified on colorbars.
Corresponding PIV analysis results are available in Movies S4, S5, S6 and S7. 
(e-h) Evolution of various parameters measured within the drops: (e) averaged divergence within the area defined by a blue circle (radius equal to half of the drop's equivalent radius), (f) averaged velocity across the entire drop, (g)  transient diffusion coefficients for PEG molecules in evaporating drops as PEG concentration increases, and (h) time-dependent Péclet numbers, calculated as $Pe = \langle |\mathbf{v}| \rangle L/D_{PEG}$.}
 \label{div}
\end{figure}

\subsection{Hydrodynamic Evolution and Its Impact on Deposit Morphology}
\label{PIVanalysis}
We applied the $\mu$-PIV technique to track the evolution of the inner flow field during the evaporation of PEG drops with molecular weights of $200$, $20\text{k}$, $300\text{k}$, and $1000\text{k}$.
These different molecular weights result in distinct deposit morphologies: spherical cap, pillar, pool-shape disk, and flat disk, respectively.
The initial PEG concentration was 2\% (w/w).

Figure~\ref{div}a-d show snapshots of the evolving 2D flow fields at a plane $\sim$\SI{25}{\micro\meter} above the substrate, corresponding to the four different molecular weights.
The red solid lines represent the contact lines at the respective time moments, while the arrows indicate the local velocity vectors $\mathbf{v}$.
The background color-coding represents the divergence of the velocity field, $\nabla \cdot \mathbf{v}$, which quantifies the local expansion or contraction of the flow.
Positive divergence (bluish tones) indicates outward flow, while negative divergence (reddish tones) reflects inward flow. 
These color maps reveal key fluid dynamics, highlighting regions of flow accumulation or depletion near the substrate.

For the drop with PEG molecular weight of $200$, the residual is a small PEG drop exhibiting a spherical-cap shape, since its bond number is $Bo = \rho_{PEG200} gL_{end}^2/\sigma_{PEG200} = 0.054$ ($L_{end} \approx \SI{0.45}{mm}$,~ $\rho_{PEG200} = $\SI{1.12e3}{kg/m^3} ~(\citet{zivkovic2013volumetric}), ~$\sigma_{PEG200} = \SI{41.5}{mN/m}$ ~(\citet{beiranvand2020experimental})).
Figure~\ref{div}a shows a flow directed from the contact line towards the drop's center, consistent with the contact line dynamics observed in CCA mode (Figs.~\ref{PEGdata}de).
To analyze this, we calculated the evolution of the averaged divergence $\langle \nabla \cdot \mathbf{v}\rangle$ within concentric circles (denoted by the blue dashed lines), with the radius set to half the temporal equivalent radius of the drop.
The evolution of this divergence, shown in Figure~\ref{div}e, reveals negative values for most of the drop's lifespan, confirming the inward flow.
A similar inward flow is observed in the evaporating drop with a PEG molecular weight of $20$k, as shown in Figure~\ref{div}b.
The calculated averaged divergence $\langle \nabla \cdot \mathbf{v}\rangle$ confirms the flow direction throughout the drying process.
However, there is a sudden spike in divergence, with a maximum occurring at around $t/t_{end}=0.9$.
This abrupt jump suggests an intense inward flow (Fig.~\ref{div}b) lasting for approximately $\sim$ \SI{30}{\second}, which coincides with the formation of the pillar (Fig.~\ref{PEGdata}b).
The underlying cause of this sharp inward flow or pillar growth remains unclear.

As the molecular weight increases, the inward flow gradually decreases and eventually reverses to an outward flow, as shown in Figure~\ref{div}c and d, respectively.
The drop with PEG molecular weight of $300k$ primarily underwent evaporation in a CCR mode (Figs.~\ref{PEGdata}de).
The pinning of the contact line, combined with a flat contact angle ($\theta<\SI{90}{\degree}$), induces outward capillary flows~(\citet{Deegan1997,larson2017twenty}), which transport liquid and polymers from the drop's center toward the pinned contact line.
This outward flow results in the formation of a pool-shaped disk, akin to the coffee ring effect.
The averaged divergence $\langle \nabla \cdot \mathbf{v}\rangle$ remains positive throughout the drop's lifespan (Fig.~\ref{div}e).
However, for the drop with a PEG molecular weight of $1000k$, indicating a very long polymer chain, the pinning effect is not sustained for the entire lifespan.
The transition from CCR to CCA mode corresponds to a shift from positive to negative divergence (Fig.~\ref{div}e).
Afterward, the system returns to the CCR mode with a pinned contact line, leading to the formation of a final flat disk.
This is not a typical coffee ring effect, likely due to the tangling of the long polymer chains~(\citet{ferry1948viscoelastic}) during the second CCR mode.

Figure~\ref{div}f compares the averaged velocity intensity for four molecular weights, showing that higher molecular weights reduce flow intensity in evaporating drops.
Inward flows dominate at lower molecular weights, while pinning and capillary flows prevail at higher molecular weights.
Increased molecular weight also alters the diffusion coefficient, affecting deposit formation~(\citet{D1,D2,D3}). 
We calculated the transient Péclet number, $Pe(t) = \langle |\mathbf{v}| \rangle L/D_{PEG}$, where $L=2R(t)$ is the contact area diameter, and $D_{PEG}(t)$ is the transient diffusion coefficient for PEG molecules as the polymer concentration increases during evaporation.
Diffusion coefficients were obtained through interpolation of data from previous studies but for dilute concentrations~(\citet{D1,D2,D3}), and Figure~\ref{div}g shows their evolution for the four cases.
Figure \ref{div}h shows that Péclet numbers for PEG $200$, $300$k, and $1000$k are similar, while PEG-$20$k has a higher Pe due to larger velocity and smaller diffusion coefficient.
Notably, during the growth of the pillar-like structure ($t/t_{end}\approx0.9$), the Péclet number shows a sharp increase, followed by a decrease once the growth is complete. 
This correlation underscores the critical role of hydrodynamics in shaping complex deposit morphologies.

\section{Conclusions}

This study provides insights into the evaporation dynamics and deposit morphologies of PEG sessile drops on hydrophobic substrates, covering a broad molecular weight range (\num{200}–\num{1000}k \unit{\gram\per\mole}).
Our results demonstrate that molecular weight significantly affects both the contact line dynamics and the internal flow patterns during evaporation, ultimately governing the resulting deposit morphologies.

At lower molecular weights (e.g., \SI{200}{\gram\per\mole}), corresponding to a liquid phase of PEG, drop evaporation favors a CCA mode, leading to the formation of spherical-cap residues.
As molecular weight increases to intermediate values (e.g., \num{20}k \unit{\gram\per\mole}), the contact line dynamics become more complex, transitioning between CCA and CCR modes, and resulting in pillar-like structures. 
At higher molecular weights (e.g., \num{70}k–\num{1000}k g/mol), evaporation stabilizes in the CCR mode, producing pool-shaped or flat disks, driven by enhanced pinning effects and capillary flows. These diverse deposit morphologies are captured in a morphological phase diagram, which highlights molecular weight as the primary determinant of deposit structure.

We employed a generalized Popov's model, incorporating time-dependent contact radius ($R(t)$) and contact angle ($\theta(t)$), to predict the mass loss rates of the drops. 
The model accurately captured evaporation behavior, with deviations at later stages attributed to non-dilute effects and the polymer solidification, confirming that vapor diffusion dominates the evaporation process across all PEG molecular weights.
Additionally, micro-particle image velocimetry ($\mu$-PIV) analysis revealed distinct internal flow patterns: inward flows for spherical caps, sharp inward flow spikes for pillars, and outward capillary flows for pool-shaped disks. 
Divergence and Péclet number analysis further linked these flow dynamics to molecular weight, with higher $Pe$ values favoring pillar formation.

Overall, these findings enhance our understanding of the thermodynamic and hydrodynamic factors that govern evaporation in polymeric sessile drops. 
The results have promising implications for controlling polymer deposition in applications such as inkjet printing and coating techniques. 
Future research could explore the effects of temperature, substrate roughness, and multicomponent systems to refine these predictive models and further enhance their practical applications.

\section*{Author Contributions}
\textbf{F.A.}: Data Curation; Formal Analysis; Investigation; Methodology; Writing – Original Draft; Writing – Review \& Editing.
\textbf{J.Y.}: Investigation; Validation; Writing – Review \& Editing.
\textbf{H.T.}: Conceptualization; Investigation; Supervision; Writing – Review \& Editing

\section*{Conflicts of interest}
There are no conflicts to declare.

\section*{Acknowledgements}
H.T. acknowledges this work is supported by the Shenzhen Fundamental Research Program (No. JCYJ20240813094626035), the National Natural Science Foundation of China (No. 12472271), 
and Guangdong Basic and Applied Basic Research Foundation (No. 2024A1515010614).

\bibliography{bibliography}
\bibliographystyle{elsarticle-num-names}

\end{document}